\documentstyle[12pt]{article}

\def\12{\frac{1}{2}}
\def\14{\frac{1}{4}}

\def\tH{\tilde{H}}
\def\tA{\tilde{A}}

\def\tf{\tilde{f}}

\def\hg{\hat{g}}
\def\hmu{\hat{\mu}}
\def\hnu{\hat{\nu}}
\def\hgam{\hat{\gamma}}
\def\hB{\hat{B}}
\def\tr{\mbox{tr}}

\begin{document}

\begin{titlepage}

\begin{flushright}
QMW-PH-98-06\\
hep-th/9802179
\end{flushright}

\vspace{3cm}

\begin{center}

{\large \bf{Geometric Actions for D-Branes and M-Branes}}

\vspace{.7cm}

M.\ Abou Zeid and C.\ M.\ Hull

\vspace{.7cm}

{\em Physics Department, Queen Mary and Westfield College, \\
Mile End Road, London E1 4NS, U.\ K.\ }

\vspace{1.5cm}

\today

\vspace{1.5cm}

\begin{abstract}

New  forms of Born-Infeld, D-brane  and M theory five-brane actions are found
which  are quadratic in the abelian field strength. The gauge fields couple
both
to a background or induced metric and a new auxiliary metric, whose  
elimination
reproduces the non-polynomial Born-Infeld action. This is similar
to the introduction of an auxiliary metric to simplify the Nambu-Goto string
action. This simplifies the  quantisation and dualisation of the gauge  
fields.

\end{abstract}

\end{center}

\end{titlepage}

\section{Introduction}

In string  theory, it has proved fruitful to replace the Nambu-Goto action
which
gives the area of the string worldsheet with a classically equivalent
action involving a worldsheet metric  and a local
conformal symmetry~\cite{BVH,DZ,HT}. The Nambu-Goto action is non-polynomial  
in
the string coordinates,
whereas the equivalent action is quadratic in the derivatives of the
coordinates, greatly simplifying the analysis and allowing a covariant
quantisation~\cite{Polya}.
This has a generalisation for the Nambu-Goto action for $p$-branes
(proportional to the world-volume), but the resulting theory is only
conformally invariant for the
string case,
$p=1$.
The purpose of this paper is to propose
and  investigate
an action that may play a similar role for the Born-Infeld theory of
electromagnetism, and its
D-brane generalisations.
The Born-Infeld action is non-polynomial in the field strength $F_{\mu \nu}$,
but introducing a
new intrinsic
 auxiliary metric
 gives a classically equivalent action which is quadratic in $F_{\mu \nu}$,
 and which has a classical conformal
symmetry in four dimensions, instead of the two dimensions for the $p$-brane
world-volume.
There are similar actions for the generalisations of Born-Infeld theory
governing the effective worldvolume theories of
D-branes~\cite{RL,FT,Li,PKT,MD,CS,GHT,Ceral,JSetal1,JSetal2,BT} and
M-branes~\cite{PST,J5,BJO,CMH1}.
As the new actions are quadratic in $F_{\mu \nu}$, integration over the gauge
fields is
straightforward and, just as in string theory,  the focus turns to the
integration over metrics.
The new action can be used to dualise the Born-Infeld
gauge field in all dimensions, circumventing the problems arising in other
approaches.
     In particular, it promises to be more convenient than the action
     presented in ref.~\cite{UL,AH} which used an auxiliary tensor field
consisting of a metric together  with an antisymmetric part.

\section{Actions}

We begin with the Born-Infeld action in $p+1$ dimensions~\cite{BIold}
\begin{equation}
S=-T_p \int d^{p+1} \sigma \sqrt{-\det (g_{\mu \nu} +F_{\mu \nu})}
\label{BI}
\end{equation}
where
\begin{equation}
F_{\mu \nu} = \partial_\mu A_\nu -\partial_\nu A_\mu
\label{FisdA}
\end{equation}
is the field strength of a $U(1)$ gauge field $A_\mu$, $\mu ,\nu =0,\ldots  
,p$
are space-time indices and $g_{\mu \nu}$ is the space-time metric. We now
show that the
action~(\ref{BI}) can be rewritten in a form which is quadratic in the
field strength $F$, and is therefore simpler to analyse and quantise. The key
is to use the
fact that
\begin{equation}
\det (g_{\mu \nu} + F_{\mu \nu} ) =
\det (g_{\mu \nu}- F_{\mu \nu})
\end{equation}
to write the integrand in~(\ref{BI}) in the form
\begin{eqnarray}
\left[ -\det (g_{\mu \nu} +F_{\mu \nu}) \right]^{\frac{1}{2}}
& = &
\left[\det (g_{\mu \nu} + F_{\mu \nu} )  \right]^{\14}
\left[\det (g_{\mu \nu}- F_{\mu \nu})    \right]^{\14} \nonumber
\\ & = & (-g)^{\14} \left\{ -\det \left[ (g_{\mu \nu} + F_{\mu \nu}) g^{\nu
\rho}
(g_{\rho \sigma} -
F_{\rho \sigma}) \right] \right\}^{\14} \nonumber \\ & = & (-g)^{\14}
\left[ -\det (g_{\mu \sigma} -g^{\nu \rho} F_{\mu \nu} F_{\rho
\sigma} )\right]^{\14} ,
\end{eqnarray}
where $g \equiv \det (g_{\mu \nu})$. The action~(\ref{BI}) can thus be
rewritten
as
\begin{equation}
S' = -T_p \int d^{p+1} \sigma (-g)^{\14} (-{\cal G} )^{\14} ,
\label{eqtoBI}
\end{equation}
where
\begin{equation}
{\cal G}_{\mu \nu} = g_{\mu \nu} -g^{\rho \sigma} F_{\mu \rho}
F_{\sigma \nu}
\end{equation}
and ${\cal G} \equiv \det [{\cal G}_{\mu \nu}]$. Introducing an
intrinsic metric $\gamma_{\mu \nu}$ allows us to rewrite~(\ref{eqtoBI}) in
the following classically equivalent form which is quadratic in the gauge  
field
strength  $F_{\mu \nu}$
\begin{eqnarray}
S' & = & -T'_p \int d^{p+1} \sigma (-g)^{\14} (-\gamma )^{\14} \left[
\gamma^{\mu \nu} {\cal G}_{\mu \nu} -(p-3) \Lambda \right] \nonumber \\ & = &
-T'_p \int
d^{p+1}\sigma (-g)^{\14} (-\gamma )^{\14} \left[ \gamma^{\mu \nu}
\left( g_{\mu \nu} -g^{\rho \sigma} F_{\mu \rho} F_{\sigma \nu} \right)
-(p-3) \Lambda \right] ,
\label{new1}
\end{eqnarray}
where $\gamma \equiv \det (\gamma_{\mu \nu})$ and $\Lambda$ is a constant.
For
$p\neq 3$, the $\gamma_{\mu \nu}$ field equation implies
\begin{equation}
\gamma_{\mu \nu} = \frac{1}{\Lambda} \left( g_{\mu \nu} -g^{\rho \sigma}
F_{\mu \rho} F_{\sigma \nu} \right)
\label{fenot3}
\end{equation}
and substituting back into~(\ref{new1}) yields the action~(\ref{eqtoBI}),
which is identical to the Born-Infeld action~(\ref{BI}). The constants $T_p$,
$T_p '$ are
related by
\begin{equation}
T_p ' =\14 \Lambda^{\frac{p-3}{4}} T_p .
\label{T's}
\end{equation}

For $p=3$, the  four-dimensional action~(\ref{new1}) is invariant under the
Weyl transformation
\begin{equation}
\gamma_{\mu \nu} \rightarrow \omega (\sigma ) \gamma_{\mu \nu}
\label{Weyl}
\end{equation}
and the $\gamma_{\mu \nu}$ field equation implies
\begin{equation}
\gamma_{\mu \nu} = \Omega {\cal G}_{\mu \nu}
\label{gfe3}
\end{equation}
for some $\Omega$, which is found by taking traces of both sides; this gives
\begin{equation}
\gamma^{\rho \sigma} \left( g_{\rho \sigma} -g^{\kappa \delta} F_{\rho  
\kappa}
F_{\delta \sigma} \right) \gamma_{\mu \nu} = 4\left( g_{\mu \nu} -g^{\rho
\sigma} F_{\mu \rho} F_{\sigma \nu} \right) .
\end{equation}
Substituting this back into~(\ref{new1}) gives~(\ref{eqtoBI}), which is
identical
to the Born-Infeld action~(\ref{BI}),
so that~(\ref{BI}) and~(\ref{new1}) are classically equivalent.

This can be generalised to the D-brane kinetic term
\begin{equation}
S=-T_p \int d^{p+1} \sigma e^{-\phi} \sqrt{ -\det ( g_{\mu \nu}
+{\cal F}_{\mu \nu} )}
\label{DBI}
\end{equation}
where
\begin{equation}
{\cal F}_{\mu \nu} \equiv F_{\mu \nu} -B_{\mu \nu},
\label{defcalF}
\end{equation}
$\phi$, $g_{\mu \nu}$ and $B_{\mu \nu}$ are the pullbacks to the
worldvolume of the background dilaton, metric and NS antisymmetric two-form
fields
and $F=dA$, with $A$
the $U(1)$ world-volume gauge field. This action gives the
effective dynamics of the zero-modes of the open strings with ends tethered
on a D-brane when
$F$ is slowly varying, so that corrections involving $\nabla F$ can be  
ignored,
and has
therefore played a central role in recent studies of
D-brane dynamics and string theory duality~\cite{JP}. However, the
non-linearity
of~(\ref{DBI}) makes
it rather difficult to study. In particular,
the action~(\ref{DBI}) is inconvenient for the purpose of quantisation,
and its dualisation has proved rather  
difficult~\cite{PKT,CS,AS,AAT,Yo,JSdual}.
It
is therefore useful
to know classically equivalent, alternative forms of this action which have
a
more tractable
dependence on the spacetime coordinates $X$ or on the
field strength $F$. In ref.~\cite{AH}, we obtained an alternative form
of~(\ref{DBI}) which is {\em linear} in $F$ and {\em quadratic} in  
derivatives
of $X$ by introducing an
auxiliary worldvolume tensor with both symmetric
and antisymmetric parts, and discussed the dualisation of the worldvolume
gauge field in this approach~\cite{AH,MA}.
Here, we give an alternative form of~(\ref{DBI}) that is {\em quadratic} in
$F$.

As before, introducing an intrinsic metric $\gamma_{\mu \nu}$ allows us to
rewrite~(\ref{DBI}) in the classically equivalent form
\begin{eqnarray}
S' & = & -T_p ' \int d^{p+1} \sigma e^{-\phi} (-g)^{\14} (-\gamma )^{\14}
\left[ \gamma^{\mu \nu} {\cal G}_{\mu \nu} -(p-3) \Lambda \right] \nonumber  
\\
& = & -T_p ' \int d^{p+1} \sigma e^{-\phi} (-g)^{\14} (-\gamma )^{\14}
\left[ \gamma^{\mu \nu} (g_{\mu \nu} -g^{\rho \sigma} {\cal F}_{\mu \rho}
{\cal F}_{\sigma \nu} ) -(p-3)\Lambda \right] ,
\label{new2}
\end{eqnarray}
where the tensions $T_p$, $T_p '$ are related as in eq.~(\ref{T's}).

The energy-momentum tensor $T_{\mu \nu}$ can be defined from the
form~(\ref{new2})
of the D-brane kinetic term by
\begin{equation}
T_{\mu \nu} \equiv  -\frac{1}{T'_{p}} \frac{1}{(-\gamma )^{\frac{1}{4}}}
\frac{\delta S}{\delta \gamma^{\mu \nu}}
\end{equation}
and we find
\begin{equation}
T_{\mu \nu}  = (-g)^{\frac{1}{4}} \left\{ -\frac{1}{4} \gamma_{\mu \nu}  
\left[
\gamma^{\rho \sigma} \left( g_{\rho \sigma} -g^{\kappa \tau} F_{\rho \kappa}
F_{\tau \sigma}
\right) -(p-3) \Lambda \right]  +g_{\mu \nu} -g^{\rho \sigma} F_{\mu \rho}
F_{\sigma \nu} \right\} .
\end{equation}
This is traceless (i.~e.\ $\gamma^{\mu \nu} T_{\mu \nu}=0$) if $p=3$ as a
result
of the Weyl
invariance~(\ref{Weyl}), and  the
equation $T_{\mu \nu}=0$ implies the field equation of the
metric~(\ref{fenot3})
or~(\ref{gfe3}).

The  low-energy effective action for an open  type I string includes the  
terms
given
by~(\ref{DBI}) with $p=9$, but with $g_{\mu \nu }, B_{\mu \nu}$ the
space-time metric and anti-symmetric
tensor gauge field
(rather than their pull-backs)~\cite{AAT}, and can be rewritten in the
equivalent
form~(\ref{new2}) with $p=9$. The dimensional reduction of the type I string
action~(\ref{DBI})
to $p+1$ dimensions gives the action for a D-$p$-brane in static gauge (and
with
vanishing RR gauge fields),
with the 9+1 vector field $A$ giving rise to a
vector and $9-p$
scalars $X_i$ on reduction. The reduction of the form~(\ref{new2}) of the
action
then gives a useful form of the static-gauge D-$p$-brane action which is
quadratic in $A,X$.

We now turn to the   reduction of~(\ref{new2}) from 9+1 to $p$+1 dimensions.
We use the notation that hatted quantities are ten-dimensional, so $\hmu =0,
\ldots ,9$,  while $\mu =0, \ldots ,p$
and $i=p+1,\ldots ,9$.
Then the vector field $A_{\hmu}=(A_\mu , X_{i})$ gives a vector and $9-p$
scalars $X_i$. We choose
  (for simplicity) a
flat space-time metric $\hg_{\hmu
\hnu}=\eta _{\hmu
\hnu}$ and vanishing 2-form $\hB_{\hmu \hnu}$, and make the following Ansatz
for the metric $\hgam_{\hmu \hnu}$:
\begin{equation}
\hgam_{\hmu \hnu} = \left( \begin{array}{cc} \gamma_{\mu \nu} +C^{i}{}_{\mu}
C^{j}{}_{\nu} \gamma_{ij} & C^{j}{}_{\mu} \gamma_{ij} \\ C^{k}{}_{\nu}
\gamma_{kj} & \gamma_{ij}
\end{array} \right) .
\label{dimred}
\end{equation}
Then the metric $\hgam_{\hmu \hnu}$ gives, as usual, a $p+1$-dimensional
metric  $\gamma_{\mu \nu}$, $9-p$ vector fields $C^{i}{}_{\mu}$ and
$(9-p)(10-p)/2$ scalar fields taking values in the coset $GL(9-p,R)/SO(9-p)$.
The inverse of~(\ref{dimred}) is
\begin{equation}
\hgam^{\hmu \hnu} = \left( \begin{array}{cc} \gamma^{\mu \nu} & - C^{\mu i}  
\\
-C^{\mu j} &
\gamma^{ij} +C^{i}{}_{\rho} \gamma^{\rho \sigma} C^{j}{}_{\sigma} \end{array}
\right)
\label{dimredinv}
\end{equation}
 and its determinant is
\begin{equation}
\det \hgam_{\hmu \hnu} = \det \gamma_{\mu \nu} \det \gamma_{ij} .
\end{equation}
Setting $F_{ij} \equiv 0$ and $F_{\mu i} \equiv \partial_\mu X_i$,
this gives the
following static gauge D-$p$-brane
action which is quadratic in both $F$ and $\partial X^i$:
\begin{eqnarray}
S'  & = &  -T_p ' \int d^{p+1} \sigma e^{-\phi}  \left[ - \det
\gamma_{\mu \nu} \det \gamma_{ij} \right]^{\14} \left[ \gamma^{\mu \nu}
(\eta_{\mu \nu}
+\eta^{ij} \partial_{\mu} X_i
\partial_{\nu} X_j
+\eta^{\rho
\sigma}  F_{\mu \rho}
F_{\nu \sigma} ) \right. \nonumber \\ & &   +2C^{\mu i} F_{\mu \rho}
\partial_\sigma X_i
\eta^{\rho \sigma}
 + (\eta^{ij} +C^{i}{}_{\upsilon} \gamma^{\upsilon
\tau}
C^{j}{}_{\tau} ) (\eta_{ij} +\eta^{\rho \sigma} \partial_\rho X_i
\partial_\sigma X_j  )
\nonumber \\ & & \left. -(p-3)\Lambda
\right] .
\label{Dpgf}
\end{eqnarray}
This quadratic action should be a convenient starting point for the study of
D-$p$-brane
dynamics, taking into account the Born-Infeld corrections.

The methods above can also be applied to the M-theory five-brane
action~\cite{PST,J5}. In the PST formulation, the kinetic part of the action  
is
\begin{eqnarray}
S & = & -T_5 \int d^6 \sigma \sqrt{- \det \left( g_{\mu \nu} +i\frac{g_{\mu \rho}
g_{\nu \lambda}}{\sqrt{-g (\partial a)^2}} \tilde{K}^{\rho \lambda} \right)}
\nonumber \\ & & -\frac{T_5}{4} \int d^6 \sigma \frac{1}{(\partial a)^2} \tilde{K}^{\mu \nu} H_{\mu \nu \rho} g^{\rho \lambda} \partial_\lambda a
\label{M5}
\end{eqnarray}
where
\begin{equation}
\tilde{K}^{\mu \nu} \equiv \frac{1}{6} \epsilon^{\mu \nu \rho \lambda \sigma
\tau} H_{\rho
\lambda \sigma} \partial_\tau a
\end{equation}
with $H_{\mu \nu \rho}=3\partial_{[\mu} b_{\nu \rho ]}$ the field stength of
the
self-dual
two-form tensor gauge field $b_{\mu \nu}$ propagating on the
world-volume, $a$ denotes the PST scalar~\cite{PST} and
\begin{equation}
(\partial a )^{2} \equiv g^{\mu \nu} \partial_\mu a \partial_\nu a .
\end{equation}
Introducing an intrinsic metric $\gamma_{\mu \nu}$ as before, the
action~(\ref{M5}) can be rewritten in the classically equivalent form
\begin{eqnarray}
S & = & -T_5 ' \int d^6 \sigma (-g)^{\14} (-\gamma )^{\14} \left[ \gamma^{\mu \nu}
\left( g_{\mu \nu} -\frac{g_{\mu \rho}g_{\nu \sigma} g_{\lambda \tau}
\tilde{K}^{\rho
\lambda} \tilde{K}^{\tau \sigma}}{g (\partial a )^{2}} \right) -2\Lambda
\right] \nonumber \\ & & -\frac{T_5}{4} \int d^6 \sigma \frac{1}{(\partial a)^2}\tilde{K}^{\mu \nu} H_{\mu \nu \rho} g^{\rho \lambda} \partial_\lambda a
\label{newM}
\end{eqnarray}
This is quadratic in the field strength $H$ and thus is more convenient for
gauge field
quantisation in the background $g_{\mu \nu}$ than~(\ref{M5}).

\section{Dual Actions}

The dualisation of the form~(\ref{new1}) of the Born-Infeld action can be
achieved
via the addition of a
Lagrange multiplier term imposing eq.~(\ref{FisdA}). Consider the action
\begin{eqnarray}
S  & =  & -T'_p \int d^{p+1} \sigma \left\{ (-g)^{\14} (-\gamma )^{\14}  
\left[
\gamma^{\mu \nu}
\left( g_{\mu \nu} -g^{\rho \sigma} F_{\mu \rho} F_{\sigma \nu} \right)
-(p-3) \Lambda \right]  \right. \nonumber \\ & & \left. + 2\tH^{\mu \nu}
\left( F_{\mu \nu} -\partial_{[\mu} A_{\nu ]} \right)
\right\},
\label{addLag}
\end{eqnarray}
where $\tH^{\mu \nu}$ is a tensor density and $F$ is regarded as an  
independent
field. Integrating out $\tH^{\mu \nu}$
sets $F=dA$ and yields  the original action~(\ref{new1}). Alternatively,
integrating out
$A_{\mu}$ imposes the constraint
\begin{equation}
\partial_\mu \tH^{\mu \nu} = 0
\label{constr}
\end{equation}
which can be solved in terms of a $(p-2)$-form $\tA$,
\begin{equation}
\tH^{\mu \nu} = \frac{1}{(p-1)!} \epsilon^{\mu \nu \rho \gamma_1 \ldots
\gamma_{p-2}} \partial_{[\rho} \tA_{\gamma_1 \ldots \gamma_{p-2}]} ,
\label{Hsol}
\end{equation}
where $\epsilon^{\mu \nu \rho \ldots}$ is the alternating tensor density. Now
$F$ is an
auxiliary two-form occuring quadratically in the action and can
be integrated out.
The field equation for $F_{\mu \nu}$ is
\begin{equation}
(-g)^{\14} (-\gamma )^{\14} \left( \gamma^{\mu \rho} g^{\nu \sigma}
+\gamma^{\nu \sigma} g^{\mu \rho} \right)
F_{\sigma \rho} =  2\tH^{\mu \nu} ,
\label{Ffe}
\end{equation}
where $\tH^{\mu \nu}$ is given by the solution~(\ref{Hsol}), and
the Gaussian integration amounts to solving this for
 $F_{\mu \nu}$ and substituting   the solution $F[ g_{\mu \nu}, \gamma_{\mu
\nu}, \tH^{\mu \nu}]$ in
the action~(\ref{new2}). This gives the dual
action $S [ g_{\mu \nu},
\gamma_{\mu \nu}, \tH^{\mu \nu}]$.
In principle, an equivalent dual action $S_D [
g_{\mu \nu} ,\tH^{\mu \nu}]$ can then be obtained by integrating out the
auxiliary metric
$\gamma_{\mu \nu}$, but in practice  this procedure is
difficult to carry out explicitly because of the non-linearity in the
worldvolume
metric of eq.~(\ref{Ffe}) and of the action $S[g_{\mu \nu},
\gamma_{\mu \nu}, \tH^{\mu \nu}]$.

Defining the matrices
\begin{equation}
f_{\mu}{}^{\nu} =F_{\mu \rho} g^{\rho \nu} ,\ \ \ \ h_{\mu}{}^{\nu}
=(-g)^{-\14}
(-\gamma )^{-\14} g_{\mu \rho} \tH^{\nu \rho} ,\ \ \ \ \beta_{\mu}{}^{\nu} =
2\left( g_{\mu \rho}\gamma^{\rho \nu} -\delta_{\mu}{}^{\nu}\right) ,
\label{defs}
\end{equation}
the equation~(\ref{Ffe}) can be written as
\begin{equation}
h=f+\{ \beta ,f \} \equiv (1+L_{\beta} ) f
\label{hLf}
\end{equation}
where for any matrices $X,Y$, the operator $L_X$ is defined by
\begin{equation}
L_X Y \equiv \{ X,Y \} .
\label{defLb}
\end{equation}
Then~(\ref{hLf}) can be inverted to give
\begin{eqnarray}
f & = & \left( 1+L_{\beta} \right)^{-1} h = \left( 1-L_{\beta} +L_{\beta}^2 -
L_{\beta}^3 +
\ldots \right) h \nonumber \\ & = & h-\{ \beta ,h \} +\{ \beta ,\{ \beta
,h \} \} -\{ \beta ,\{ \beta ,\{ \beta ,h \} \} \}  +\ldots .
\label{series}
\end{eqnarray}
Substituting this solution for $F$ back in~(\ref{addLag}) gives
\begin{eqnarray}
S  & = & -T'_p \int d^{p+1} \sigma \left\{ (-g)^{\14} (-\gamma )^{\14}
\left[ \gamma^{\mu \nu} g_{\mu \nu} -(p-3) \Lambda \right] \right. \nonumber  
\\
& & \left.
+2(-g)^{\14} (-\gamma )^{\frac{1}{4}}  \tr \left[  h (1+L_{\beta})^{-1}
h
\right]  \right\} \nonumber \\ & = & -T'_p \int d^{p+1} \sigma \left\{
(-g)^{\14}
(-\gamma )^{\14}
\left[ \gamma^{\mu \nu} g_{\mu \nu} -(p-3) \Lambda \right]  \right. \nonumber
\\
& & \left.
+2(-g)^{-\frac{1}{2}} (-\gamma )^{-\frac{1}{2}} \tH^{\mu \sigma}
M_{\mu
\rho \nu \sigma} \tH^{\nu \rho} \right\},
\label{Sdual}
\end{eqnarray}
where the tensor $M_{\mu \nu \rho \sigma}$ is
defined by
\begin{equation}
\tr \left[  h (1+L_{\beta})^{-1} h
\right]= h^{\mu \sigma} M_{\mu
\rho \nu \sigma} h^{\nu \rho}
\end{equation}
(where $h^{\mu \sigma}=g^{\mu \tau}h_\tau {}^\sigma$)
and is given to lowest orders by
\begin{eqnarray}
M_{\mu \nu \rho \sigma} & = & \gamma_{\mu \kappa} g_{\nu \tau} \left[
\delta^{\kappa}_{\rho}
\delta^{\tau}_{\sigma} -\Sigma^{\tau}{}_{\rho}
\Sigma_{\sigma}{}^{\kappa} +\Sigma^{\alpha}{}_{\rho} \Sigma^{\tau}{}_{\alpha}
\Sigma_{\sigma}{}^{\beta} \Sigma_{\beta}{}^{\kappa}   
-\Sigma^{\alpha}{}_{\rho}
\Sigma^{\beta}{}_{\alpha} \Sigma^{\tau}{}_{\beta} \Sigma_{\sigma}{}^{\delta}
\Sigma_{\delta}{}^{\kappa} \right. \nonumber \\ & & \left.
+\Sigma^{\alpha}{}_{\rho}
\Sigma^{\beta}{}_{\alpha} \Sigma^{\delta}{}_{\beta} \Sigma^{\tau}{}_{\delta}
\Sigma_{\sigma}{}^{\epsilon} \Sigma_{\epsilon}{}^{\lambda}
\Sigma_{\lambda}{}^{\kappa} +\ldots
\right] ,
\label{evalM}
\end{eqnarray}
where
\begin{equation}
\Sigma^{\mu}{}_{\nu} \equiv g_{\nu \rho}\gamma^{\rho \mu}
\end{equation}
and $\Sigma_{\mu}{}^{\rho}$ denotes the inverse of the matrix
$\Sigma^{\mu}{}_{\rho}$.
The auxiliary metric $\gamma _{\mu \nu}$ occurs algebraically and can in
principle be eliminated
using its equation of motion, giving $\gamma _{\mu \nu}$ as a function of
$g_{\nu \rho}$ and $\tilde H^{\mu \nu}$. Although this is hard to do explicitly, it can be
done
perturbatively, giving $\gamma _{\mu \nu}$ to any desired order in $\tilde
H^{\mu \nu}$.

The dualisation of the action~(\ref{new2}), which is classically equivalent
to the D-brane kinetic term~(\ref{DBI}),  proceeds in a
similar way. Consider the action
\begin{eqnarray}
S & = & -T_p ' \int d^{p+1} \sigma \left\{ e^{-\phi} (-g)^{\frac{1}{4}}
(-\gamma
)^{\frac{1}{4}} \left[
\gamma^{\mu \nu} \left( g_{\mu \nu} -g^{\rho \sigma} {\cal F}_{\mu \rho}  
{\cal
F}_{\sigma \nu} \right) -
(p-3) \Lambda \right] \right. \nonumber \\ & & \left. + 2\tH^{\mu \nu} \left(
F_{\mu \nu}
-2\partial_{[\mu}A_{\nu ]} \right) \right\}  .
\end{eqnarray}
Integrating out $\tH^{\mu \nu}$
yields the original action~(\ref{new2}). Alternatively, integrating over
$A_\mu$
imposes the constraint~(\ref{constr}), which is solved in terms of a $(p-2)$
form $\tA$ as
in~(\ref{Hsol}). Now $F$ is an auxiliary two-form
occuring algebraically. The field equation for $F_{\mu \nu}$ is
\begin{equation}
 (-g)^{\frac{1}{4}} (-\gamma )^{\14} \left(
\gamma^{\mu \rho} g^{\nu \sigma} +\gamma^{\nu \sigma}
g^{\mu \rho} \right) (F_{\sigma \rho} -B_{\sigma \rho} ) = 2\tH^{\mu \nu}  ,
\label{Lfe}
\end{equation}
where $\tH^{\mu \nu}$ is given by the solution~(\ref{Hsol}).

Defining the matrix
\begin{equation}
\tilde{f}_{\mu}{}^{\nu} \equiv \left( F_{\mu \rho}-B_{\mu \rho} \right)  
g^{\rho
\sigma}
\end{equation}
the equation~(\ref{Lfe}) can be written as
\begin{equation}
h= \tf +\{ \beta, \tf \} = (1+L_\beta ) \tf
\end{equation}
where the matrices $h$, $\beta$ and the operator $L_\beta$ are defined as
in~(\ref{defs})
and~(\ref{defLb}). This can be inverted to give
\begin{equation}
 \tf =(1+L_\beta )^{-1} h= h-\{ \beta ,h \} +\{ \beta ,\{ \beta
,h \} \} -\{ \beta ,\{ \beta ,\{ \beta ,h \} \} \}  +\ldots .
\end{equation}
Substituting this solution for $\cal F$ back in~(\ref{addLag}) gives
\begin{eqnarray}
S  & = & -T'_p \int d^{p+1} \sigma \left\{ (-g)^{\14}
(-\gamma )^{\frac{1}{4}}
\left[ \gamma^{\mu \nu} g_{\mu \nu} -(p-3) \Lambda \right] +2\tH^{\mu
\nu}B_{\mu
\nu} \nonumber
\right. \\ & & \left.  +2(-g)^{-\frac{1}{2}} (-\gamma )^{-\frac{1}{2}}  
\tH^{\mu
\sigma} M_{\mu
\rho \nu \sigma} \tH^{\nu \rho} \right\},
\end{eqnarray}
with the tensor $M_{\mu \nu \rho \sigma}$  defined as in~(\ref{evalM}).

\section{Conclusion}

In this paper, we have presented new forms of Born-Infeld as well as D-brane
and
M
theory five-brane kinetic terms which
are quadratic in the abelian gauge field strength. The gauge fields couple  
both
to a
background
or induced metric $g_{\mu \nu}$ and to a new intrinsic metric $\gamma_{\mu
\nu}$,
and both of these world-volume
metrics appear in the action in a remarkably symmetric way. These actions  
could
play an important role in the
quantisation of Born-Infeld theory and of  the static gauge effective
world-volume theories of
D-Branes and M-Branes, similar to the role played in string theory by the
actions
of ref.~\cite{BVH,DZ,HT,Polya}.

The dualisation of the $U(1)$ gauge fields is achieved by adding a Lagrange
term
imposing the
constraint~(\ref{FisdA}), and the dual action is quadratic in the field
strength
of the
appropriate dual potential. The dual action involves an infinite power series
in
the auxiliary
intrinsic metric, which can be eliminated perturbatively. 

The four dimensional action~(\ref{new1}) with $p=3$ has a classical Weyl
invariance~(\ref{Weyl}), which is closely related to that of the string.
Quantum mechanically, this will be anomalous~\cite{CD1,CD2}.
We hope to return to  a discussion of this anomaly elsewhere, but we note  
here
that it is
trivial to generalise our action to that for a theory of $N$ abelian vector
fields and $Nd$ scalars $X_i$, and it is
intriguing that it may be possible to choose the numbers $N,d$ to take  
critical
values that give  a
cancellation of the conformal anomaly, generalising the critical dimension of
string theory.

\end{document}